\documentclass[twocolumn,showpacks]{revtex4}
\usepackage[english]{babel}
\usepackage{graphicx,amsmath,amsfonts,amssymb,amsbsy}
\usepackage[dvips]{color}
\usepackage{epsfig}
\usepackage{natbib}
\usepackage{pstricks}
\usepackage{mathrsfs}
\usepackage{color}
\usepackage{soul}
\usepackage[normalem]{ulem}

\newcommand{\msun}{M_{\odot}}

\begin{document}

\title{Thermal evolution of hybrid stars within the framework of a
  non-local NJL model}

\author{S.~M.~de~Carvalho$^{1,2}$, R.~Negreiros$^{2}$, M.~Orsaria$^{3,4}$,
G.~A.~Contrera$^{3,4,5}$, F.~Weber$^{6,7}$ and W. Spinella$^{6,8}$}
\affiliation{$^1$ICRANet-Rio, Centro Brasileiro de Pesquisas F\'isicas, Rua Dr.
Xavier Sigaud 150, Rio de Janeiro, RJ, 22290-180, Brazil }
\affiliation{$^2$Instituto de F\'isica, Universidade Federal Fluminense, UFF,
Niter\'oi, 24210-346, RJ, Brazil}
\affiliation{$^3$CONICET, Rivadavia 1917, 1033 Buenos Aires, Argentina}
\affiliation{$^4$Grupo de Gravitaci\'on, Astrof\'isica y Cosmolog\'ia,\\
Facultad de Ciencias Astron{\'o}micas y
  Geof{\'i}sicas, Universidad Nacional de La Plata,\\ Paseo del Bosque S/N
(1900), La Plata, Argentina}
\affiliation{$^5$IFLP, CONICET - Dpto. de F{\'i}sica, UNLP, La Plata, Argentina}
\affiliation{$^6$Department of Physics, San Diego State University, 5500
    Campanile Drive, San Diego, California 92182}
\affiliation{$^7$Center for Astrophysics
    and Space Sciences, University of California,\\ San Diego, La Jolla,
    CA 92093, USA}
\affiliation{$^8$Computational Science Research Center, San Diego State
University, 5500 Campanile Drive, San Diego, California 92182}

\begin{abstract}
We study the thermal evolution of neutron stars containing deconfined
quark matter in their core. Such objects are generally referred to as
quark-hybrid stars. The confined hadronic matter in their core is
described in the framework of non-linear relativistic nuclear field
theory. For the quark phase we use a non-local extension of the SU(3)
Nambu Jona-Lasinio model with vector interactions. The Gibbs condition
is used to model phase equilibrium between confined hadronic matter
and deconfined quark matter.  Our study indicates that high-mass
neutron stars may contain between 35 and 40\% deconfined quark-hybrid
matter in their cores. Neutron stars with canonical masses of around
$1.4\, M_\odot$ would not contain deconfined quark matter. The central
proton fractions of the stars are found to be high, enabling them to
cool rapidly.  Very good agreement with the temperature evolution
established for the neutron star in Cassiopeia A (Cas A) is obtained
for one of our models (based on the popular NL3 nuclear
parametrization), if the protons in the core of our stellar models are
strongly paired, the repulsion among the quarks is mildly repulsive,
and the mass of Cas A has a canonical value of $1.4\, M_\odot$.
\end{abstract}

\date{\today}



\maketitle

\section{Introduction}
\label{sect:1}

Exploring the properties of compressed baryonic matter, or, more
generally, strongly interacting matter at high densities and/or
temperatures, has become a forefront area of modern physics.
Experimentally, the properties of such matter are being probed with
the Relativistic Heavy Ion Collider RHIC at Brookhaven and the Large
Hadron Collider LHC at Cern. Great advances in our understanding of
such matter are also expected from the next generation of heavy-ion
collision experiments at FAIR (Facility for Antiproton and Ion
Research at GSI) and NICA (Nucloton-based Ion Collider fAcility at
JINR) \cite{CBMbook11:a,NICA} as well as from the study of neutron
stars (for an overview, see \cite{glendenning00:book,fridolin1,%
  blaschke01:trento,lattimer01:a,weber05:a,%
  page06:review,haensel06:book,klahn06:short,sedrakian07:a,klahn07:a,%
  alford08:a,CBMbook11:a,Weber:2014qoa,grigorian15:a} and references
therein).

Neutron stars (NSs) contain nuclear matter compressed to densities
which are several times higher than the densities of atomic nuclei.
At such extreme conditions, the fundamental building blocks of matter
may no longer be just neutrons and protons immersed in a gas of
relativistic electron and muons, but other nuclear degrees of freedom
such as hyperons, delta particles and, most intriguingly, deconfined
up, down and strange quarks may begin to play a role.  Neutron stars
containing deconfined quark matter in their central core are referred
to as quark-hybrid stars (hybrid stars, for short).

The most massive neutron stars observed to date are
  $J1614-2230$ ($1.97 \pm 0.04 \, M_{\odot}$)~\cite{Demorest2010} and
  $J0348+0432$ ($2.01 \pm 0.04\, M_{\odot}$)~\cite{Antoniadis13}.  In
  several recent papers
  \cite{Orsaria:2013,Orsaria:2014,Chen:2013tfa,Chen:2015mda,%
    Alvarez-Castillo:2014dva,Lastowiecki:2015mpa,Dexheimer:2014pea},
  it has been shown that they may contain significant fractions of
  quark-hybrid matter in their centers, despite the relatively stiff
  nuclear equation of state (EoS) that is required to achieve such
  high masses. The radii of these neutron stars would be between 13
  and 14~km, depending on the nuclear EoS
  \cite{Orsaria:2013,Orsaria:2014}, increasing to respectively 13.5
  and 14.5~km for lighter neutron stars with canonical masses of
  around $1.4\, M_\odot$. Such radius values lie between the radius
  determinations based on X-ray burst oscillations of neutron stars in
  low-mass X-ray binaries \cite{Ozel:2015,Guillot:2014,Steiner:2013,Suleimanov:2011}
  and the estimates of the radius of the isolated neutron star RX
  J1856-3754 \cite{Truemper:2011}, emitting purely thermal radiation
  in the X-ray and in the optical bands.

If the dense interior of a neutron star contains deconfined quark
matter, it will most likely be three-flavor quark matter, since such
matter has lower energy than two-flavor quark matter
\cite{ellis91:a,glendenning92:a}. Furthermore, just as for the hyperon
content of neutron stars, strangeness is not conserved on macroscopic
time scales which allows neutron stars to convert confined hadronic
matter to three-flavor quark matter until equilibrium brings this
process to a halt.  We considered the transition from hadronic to
quark matter to be first order. There are two distinct ways to
construct a first order phase transition in neutron stars. The first
option is a Gibbs construction, where the electronic and baryonic
chemical potentials as well as the pressure are continuous during the
phase transition; the second option is a Maxwell construction, where
only the baryonic chemical potential and pressure are continuous and
the electronic chemical potential is characterized by a discontinuity
at the phase boundary. The surface tension at the interface between
the quark-hadron phase is what determines whether a Gibbs or Maxwell
phase transition may be taking place. Several authors
\cite{lida:2008,Palhares:2010be,endo:2011,Pinto:2012aq,lugones:2013,Ke:2013wga}
have attempted to estimate the value of the surface tension,  with
mixed results. For what follows, we will assume that the surface
tension is less than 40~MeV~fm$^{-2}$, such that the Gibbs condition
is favored and a mixed phase of quark matter and nuclear matter exists
above a certain density \cite{alford01:b}.

In the mixed phase, the presence of quarks allows the hadronic
component to become more isospin symmetric, which is accomplished by
the transference of electric charge to the quark phase. Thus, the
symmetry energy can be lowered at only a small cost in rearranging the
quark Fermi surfaces. The implication of this charge rearrangement is
that the mixed phase region of a neutron star will have positively
charged hadronic matter and negatively charged quark
matter~\cite{glendenning00:book,glendenning92:a,glendenning01:a}. This
should have important implications for the electric and thermal
properties of NSs.  Studies of the transport properties of
quark-hybrid neutron star matter have been reported in
~\cite{reddy00:a,na12:a}.

As already mentioned above, this study is carried out for neutron
stars containing deconfined quark matter in their centers (so-called
quark-hybrid stars). To describe the quark matter phase, we use a
non-local extension of the SU(3) Nambu Jona-Lasinio (NJL) model
\cite{Nambu,Scarpettini,Contrera2008,Contrera2_2010} with vector
interactions. For the hadronic phase we consider a non-linear
relativistic mean-field model
\cite{Walecka1974,boguta77:a,boguta77:b,boguta83:a,Serot1986} solved
for two different parametrizations, GM1~\cite{Glendenning1991} and
NL3~\cite{Lalazissis}. The transition from the confined hadronic phase
to the deconfined quark phase is treated as a Gibbs transition,
imposing global electric charge neutrality and baryon number
conservation on the field equations. We find that the non-local NJL
model predicts the existence of extended regions of mixed quark-hadron
(quark-hybrid) matter in neutron stars with masses up to $2.4\,
M_{\odot}$.

The paper is organized as follows. In Sect.\ \ref{sect:2}, we describe
the non-local extension of the SU(3) NJL model at zero temperature. In
Sect.\ \ref{sect:3}, the non-linear relativistic mean-field model,
which is used for the description of confined hadronic matter, is
briefly discussed.  In Sect.\ \ref{sect:4}, the construction of the
quark-hadron mixed phase is discussed for neutron star matter
characterized by global charge neutrality. Our results for the global
structure and composition of quark-hybrid stars are discussed in
Sects.\ \ref{sect:5}. A discussion of their thermal evolution is
presented in \ref{sect:6}. Finally, a summary of our results and
important conclusions are provided in Sect.\ \ref{sect:7}. \\

\section{Quark Matter Phase}\label{sect:2}

In this section we briefly describe the non-local extension of the
SU(3) Nambu Jona-Lasinio (n3NJL) model.  The Euclidean effective
action for the quark sector, including the vector coupling
interaction, is given by
\begin{eqnarray}
S_E &=& \int d^4x \left\{ \bar \psi (x) \left[ -i \partial{\hskip-2.0mm}/ +
\hat m \right] \psi(x) \right.
\nonumber \\ &-&
 \frac{G_S}{2} \left[ j_a^S(x) \ j_a^S(x) + j_a^P(x) \ j_a^P(x) \right]
\nonumber \\ &-&
 \frac{G_P}{4} \ T_{abc} \left[ j_a^S(x)
  j_b^S(x) j_c^S(x) - 3\ j_a^S(x) j_b^P(x) j_c^P(x) \right]
\nonumber \\ &-&
\left.   \frac{G_{V}}{2} \left[j^\mu_{V a}(x)
j^\mu_{V a}(x)\right]  \right\} \, ,
\label{se}
\end{eqnarray}
where $\psi$ stands for the light quark fields, $\hat m$ denotes the
current quark mass matrix, and $G_S$, $G_P$ and $G_V$ are the scalar,
pseudo-scalar, and vector coupling constant of the theory,
respectively. For simplicity, we consider the isospin symmetric limit
in which $m_u = m_d=\bar m$.  The operator $\partial{\hskip-2.0mm}/ =
\gamma_\mu\partial_\mu$ in Euclidean space is defined as $\vec \gamma
\cdot \vec \nabla + \gamma_4\frac{\partial}{\partial \tau}$, with
$\gamma_4=i\gamma_0$. The scalar ($S$), pseudo-scalar ($P$), and
vector ($V$) current densities $j_a^{S}(x)$, $j_a^{P}(x)$, and $j_{V
  a}^{\mu}(x)$, respectively, are given by
\begin{align}
j_{a}^S(x) & =\int d^{4}z\ \widetilde{g}(z)\ \bar{\psi}\left(
x+\frac{z}{2}\right) \ \lambda_{a}\ \psi\left( x-\frac{z}{2}\right)
\ ,\nonumber\\ j_{a}^P(x) & =\int
d^{4}z\ \widetilde{g}(z)\ \bar{\psi}\left( x+\frac{z}{2}\right) \ i
\ \gamma_5 \lambda_{a} \ \psi\left(
x-\frac{z}{2}\right)\ ,\nonumber\\ j^{\mu}_{V a}(x) & =\int
d^{4}z\ \widetilde{g}(z)\ \bar{\psi}\left( x+\frac{z}{2}\right)
\ \gamma^{\mu}\lambda_{a}\ \psi\left( x-\frac{z}{2}\right) \, ,
\label{currents}
\end{align}
where $\widetilde{g}(z)$ is a form factor responsible for the
non-local character of the interaction, $\lambda_a$ with $a=1,
\ldots,8$ denotes the generators of SU(3), and $\lambda_0=\sqrt{2/3}\,
\openone_{3\times 3}$.  Finally, the constants $T_{abc}$ in the t'Hooft
term accounting for flavor-mixing are defined by
\begin{equation}
T_{abc} = \frac{1}{3!} \ \epsilon_{ijk} \ \epsilon_{mnl} \
\left(\lambda_a\right)_{im} \left(\lambda_b\right)_{jn}
\left(\lambda_c\right)_{kl}\;.
\end{equation}

After the standard bosonization of Eq.\ (\ref{se}), the integrals over
the quark fields can be performed in the framework of the Euclidean
four-momentum formalism.  Thus, the grand canonical thermodynamical
potential of the model within the mean field approximation at zero
temperature is given by
\begin{widetext}
\begin{eqnarray}
\label{gcp}
&&\Omega^{NL} (M_f,\mu_f)  =  -\frac{N_c}{\pi^3}\sum_{f=u,d,s}
\int^{\infty}_{0} dp_0 \int^{\infty}_{0}\, dp \,\mbox{ ln }\left\{
\left[\widehat \omega_f^2 + M_{f}^2(\omega_f^2)\right]
\frac{1}{\omega_f^2 + m_{f}^2}\right\} \label{omzerot} \\ & & -
\frac{N_c}{\pi^2} \sum_{f=u,d,s} \int^{\sqrt{\mu_f^2-m_{f}^2}}_{0}
dp\,\, p^2\,\, \left[(\mu_f-E_f) \Theta(\mu_f-m_f) \right]
- \frac{1}{2}\left[\sum_{f=u,d,s} (\bar \sigma_f \ \bar S_f +
  \frac{G_S}{2} \ \bar S_f^2) + \frac{G_P}{2} \bar S_u\ \bar S_d\ \bar
  S_s\right]
- \sum_{f=u,d,s}\frac{\varpi_f^2}{4 G_V} \, , \nonumber
\end{eqnarray}
\end{widetext}
where $N_c=3$, $E_{f} = \sqrt{p^{2} + m_{f}^{2}}$, $\omega_f^2 =
(\,p_0\,+\,i\,\mu_f\,)^2\,+\,p^2$, and $\bar\sigma_f$ denotes the
mean-field values of the quark flavor ($f=u,d,s$) fields.  The vector
coupling constant $G_V$ is treated as a free parameter and expressed
as a fraction of the strong coupling constant $G_S$.

The constituent quark masses $M_{f}$ are treated as momentum-dependent
quantities. They are given by
\begin{equation}
M_{f}(\omega_{f}^2) \ = \ m_f + \bar\sigma_f g(\omega_{f}^2)\, ,
\end{equation}
where $g(\omega^2_f)$ is the Fourier transform of the form factor
$\widetilde{g}(z)$.  The vector mean fields $\varpi_f$ are associated
with the vector current densities $j_{V a}^{\mu}(x)$, where a
different vector field for each quark flavor $f$ has been considered.

We followed the method described in \cite{Blaschke:2007ri} to include
the vector interactions. The inclusion of vector interactions shifts
the quark chemical potential as follows,
\begin{eqnarray}
 \widehat{\mu}_f &=&\mu_f - g(\omega^2_f)\varpi_f\, , \\
\widehat{\omega}_f^2 &=& (\,p_0\, + \,i\,\widehat{\mu}_f\,)^2\, +
\,p^2 \, .
\end{eqnarray}
Note that the shift in the quark chemical potential does not affect
the Gaussian non-local form factor,
\begin{equation}
g(\omega^2_f) = \exp{\left(-\omega^2_f/\Lambda^2\right)}\, ,
\label{rg}
\end{equation}
avoiding a recursive problem as discussed in
\cite{Blaschke:2007ri,Weise2011,Contrera:2012wj}.  In Eq.\ (\ref{rg}),
$\Lambda$ plays the role of an ultraviolet cut-off momentum scale and
is taken as a parameter which, together with the quark current masses
and coupling constants $G_S$ and $G_P$ in Eq.\ (\ref{se}), can be
chosen so as to reproduce the phenomenological values of pion decay
constant $f_\pi$, and the meson masses $m_{\pi}$, $m_\eta$,
$m_{\eta'}$, as described in \cite{Contrera2008,Contrera2_2010}.  In
this work we use the same parameters as in
\cite{Orsaria:2013,Orsaria:2014}.

For the stationary phase approximation \cite{Scarpettini}, the
mean-field values of the auxiliary fields $\bar S_f$ in
Eq.\ (\ref{gcp}) are given by
\begin{equation}
\bar S_f = -\, 4\,N_c\, \int^{\infty}_{0}\,\frac{dp_0}{2\pi}
\int \frac{d^3p}{(2\pi)^3} \, g(\omega_f^2)\, \frac{
  M_{f}(\omega_f^2)}{\widehat{\omega}^2 + M_{f}^2(\omega_f^2)}\, .
\end{equation}
Due to the charge neutrality constraint, we consider three scalar
fields, $\bar \sigma_u$, $\bar \sigma_d$ and $\bar \sigma_s$, which can
be obtained by solving the coupled system of ``gap'' equations
\cite{Scarpettini} given by
\begin{eqnarray}
\bar \sigma_u + G_S\,\bar
S_u + \frac{G_P}{2} \, \bar S_d \bar S_s &=& 0\, , \nonumber \\
\bar \sigma_d + G_S\,\bar
S_d + \frac{G_P}{2} \, \bar S_u \bar S_s &=& 0\, , \\
 \bar \sigma_s + G_S\,\bar
S_s + \frac{G_P}{2} \, \bar S_u \bar S_d &=& 0\, . \nonumber
\end{eqnarray}
The vector mean fields $\varpi_{f}$ are obtained by minimizing
Eq.\ (\ref{gcp}), i.e.  $\frac{\partial \Omega^{\rm NL}}{\partial
  {\varpi_{f}}}= 0$.

For quark matter in chemical equilibrium at finite density, the basic
particle processes involved are given by the strong process $u+d
\leftrightarrow u+s$ as well as the weak processes $d(s)\rightarrow u
+ e^{-}$ and $u+e^{-}\rightarrow d(s)$.  We are assuming that
neutrinos, once created by weak reactions, leave the system, which is
equivalent to supposing that the neutrino chemical potential is equal
to zero. Therefore, the chemical potential for each quark flavor $f$
is given by
\begin{equation}
\mu_f = \mu_b - Q \mu_e \, ,
\end{equation}
where $Q = {\rm diag}(2/3,-1/3,-1/3)$ in flavor space and $\mu_b = 1/3
\sum_f \mu_f$ is the baryonic chemical potential.

The contribution of free degenerate leptons to the quark phase is
given by
\begin{equation}
\label{elezero}
\Omega_{\lambda=e^{-} , \mu^{-}}(\mu_e) = - \frac{1}{\pi^2}
\int_{0}^{p_{F_{\lambda}}} p^2 \left(\sqrt{p^2 + m_{\lambda}^{2} } -
\mu_e \right) dp \, .
\end{equation}
Muons appear in the system if the electron chemical potential $\mu_e =
\mu_\mu$ is greater than the muon rest mass, $m_\mu = 105.7$~MeV. For
electrons we have $m_e = 0.5$ MeV. Thus, the total thermodynamic
potential of the quark phase is given by Eq.\ (\ref{omzerot})
supplemented with the leptonic contribution of Eq.\ (\ref{elezero}).

\section{Confined Hadronic Matter}
\label{sect:3}

The hadronic phase is described in the framework of the non-linear
relativistic nuclear field theory
\cite{Walecka1974,boguta77:a,boguta77:b,boguta83:a,Serot1986}, where
baryons (neutrons, protons, hyperons and delta states) interact via
the exchange of scalar, vector and isovector mesons ($\sigma$,
$\omega$, $\rho$, respectively). The parametrizations used in our
study are GM1 \cite{Glendenning1991} and NL3 \cite{Lalazissis}.
The lagrangian of this model is given by
\begin{eqnarray}
\mathcal{L} = \mathcal{L}_{\it H} + \mathcal{L}_{\ell} \, ,
\label{eq:lag}
\end{eqnarray}
with the leptonic lagrangian given by
\begin{eqnarray}
\mathcal{L}_{\ell} = \sum_{\lambda=e^-, \mu^-} \bar{\psi}_\lambda
(i\gamma_\mu\partial^\mu - m_\lambda) \psi_\lambda \, .
\label{eq:leplag}
\end{eqnarray}
The hadronic lagrangian has the form
\begin{widetext}
\begin{eqnarray}
  \mathcal{L}_{\it H} &=& \sum_{B=n,p, \Lambda, \Sigma,
    \Xi}\bar{\psi}_B \bigl[\gamma_\mu (i\partial^\mu - g_\omega
    \omega^\mu - g_\rho \vec{\rho}_\mu) - (m_N - g_\sigma\sigma)
    \bigr] \psi_B + \frac{1}{2} (\partial_\mu \sigma\partial^\mu
  \sigma - m_\sigma^2 \sigma^2) - \frac{1}{3} b_\sigma m_N (g_\sigma
  \sigma)^3 - \frac{1}{4} c_\sigma (g_\sigma \sigma)^4 \nonumber \\
&&- \frac{1}{4}\omega_{\mu\nu} \omega^{\mu\nu}
  +\frac{1}{2}m_\omega^2\omega_\mu \omega^\mu + \frac{1}{2}m_\rho^2
  \vec{\rho}_\mu \cdot \vec{\rho\,}^\mu - \frac{1}{4}
  \vec{\rho}_{\mu\nu} \vec{\rho\,}^{\mu\nu} . \label{eq:Blag}
\end{eqnarray}
\end{widetext}
The quantity $B$ sums over all baryonic states which become populated
in neutron star matter at a given density
\cite{glendenning00:book,fridolin1}. Intriguingly, we have found
 that, aside from hyperons, the $\Delta^-$ state
becomes populated in neutron star matter at densities that could be
reached in the cores of stable neutron stars \cite{Orsaria:2014}.  Simpler treatments of
the quark-hadron phase transition, based on the MIT bag model
\cite{glendenning92:a,glendenning01:a} do not predict the occurrence
of the $\Delta^-$ state in stable neutron stars.

For the models of this paper, $\Delta$ states become populated when
the vector repulsion among quarks reaches values of $G_V \gtrsim 0.05
\, G_S$, leading to a substantial stiffening of the EoS. This
stiffening more than offsets the softening of the EoS caused the
generation of the $\Delta$ states, resulting in an EoS which is
readily capable to accommodate even very heavy ($2 \, M_\odot$)
neutron stars. Without this additional stiffening, it would be
difficult to account for $2 \, M_\odot$ neutron stars with equations
of state that are characterized by an early appearance of $\Delta$'s,
at densities between around 2 to 3 times nuclear saturation density
\cite{Drago:2014}.

\section{Quark-Hadron Mixed Phase}
\label{sect:4}

To determine the mixed phase region of quarks and hadrons we start
from the Gibbs condition for pressure equilibrium between confined
hadronic ($P^H$) matter and deconfined quark ($P^q$) matter.  The
Gibbs condition is given by
\begin{eqnarray}
 P^H(\mu_b^H, \mu_e^H, \{\phi \} ) = P^q(\mu_b^q, \mu_e^q, \{\psi \} )
 \, ,
\label{eq:GibbsP}
\end{eqnarray}
with $\mu_b^H = \mu_b^q$ for the baryon chemical potentials and
$\mu_e^H=\mu_e^q$ for the electron chemical potentials in the hadronic
($H$) and quark ($q$) phase. The quantities $\{ \phi \}$ and $\{
\psi\}$ stand collectively for the field variables and Fermi momenta
that characterize the solutions to the equations of confined hadronic
matter and deconfined quark matter, respectively. By definition, the
quark chemical potential is given by $\mu_b^q = \mu_n/3$, where
$\mu_n$ is the chemical potential of the neutrons.  In the mixed
phase, the baryon number density, $n_b$, and the energy density,
$\varepsilon$, are given by
\begin{equation}
    n_b = (1-\chi) n_b^H + \chi n_b^q \, ,
\end{equation}
and
\begin{equation}
    \varepsilon = (1-\chi) \varepsilon^H + \chi\varepsilon^q\, ,
\end{equation}
where $n_b^H$ ($\varepsilon^H$) and $n_b^q$ ($\varepsilon^q$) denote
the baryon number (energy) densities of the hadronic and the quark
phase, respectively. The quantity $\chi \equiv V_q/V$ denotes the
volume proportion of quark matter, $V_q$, in the unknown volume
$V$. Therefore, by definition $\chi$ varies from 0 to 1 depending
on how much confined hadronic matter has been converted to quark
matter at a given density.
\begin{figure}[!hbtp]
\includegraphics[width=.9\linewidth,clip]{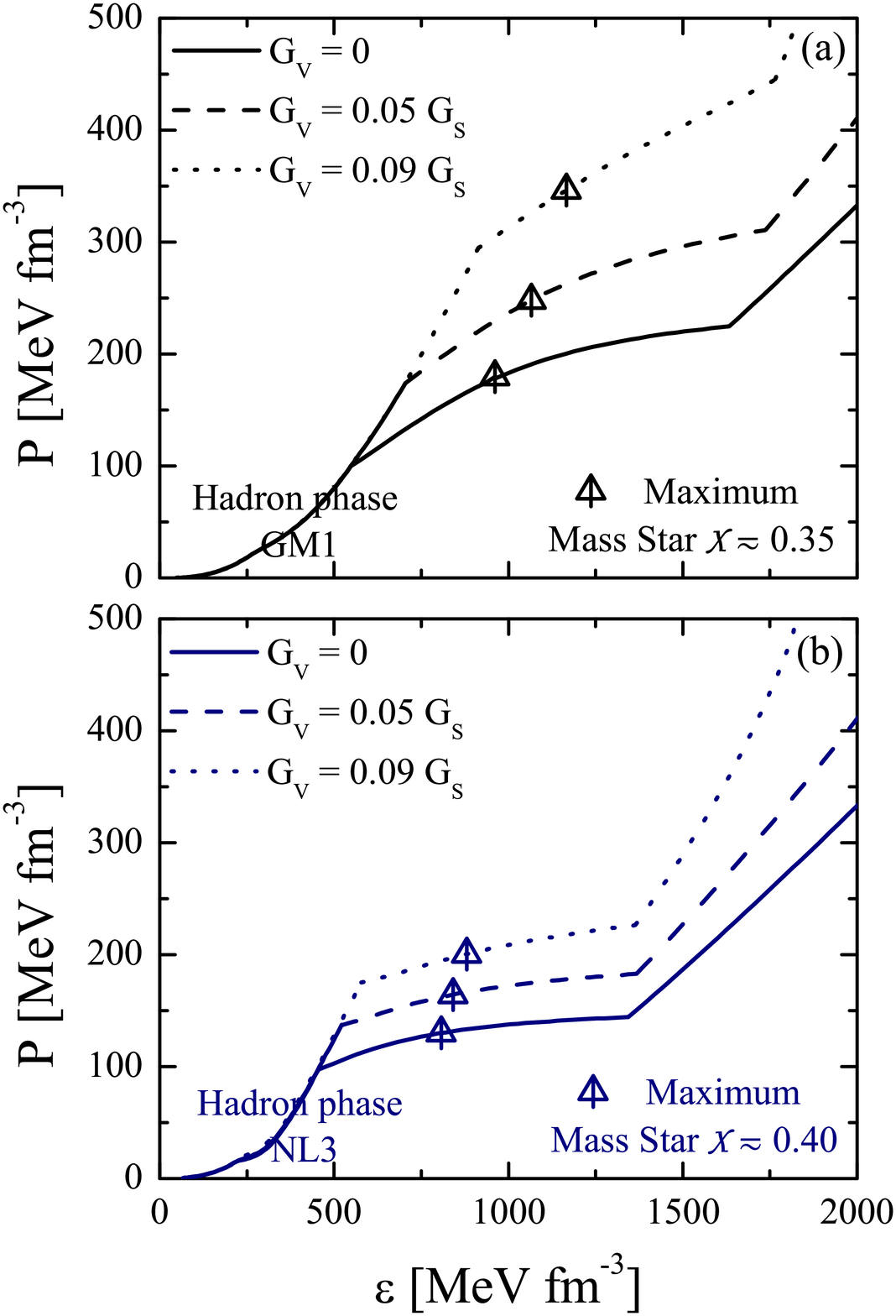}
\caption{(Color online) Pressure, $P$, as a function of the energy
  density, $\epsilon$, for the different nuclear parametrizations
  (GM1, NL3) and vector coupling constant $G_V/G_S$ $(0, 0.05, 0.09)$
  considered in this paper.  Panel (a) shows the hybrid EoSs computed
  for GM1, panel (b) shows the hybrid EoSs computed for NL3.  The
  triangles in both panels indicate the central densities of the
  maximum-mass neutron stars (see Fig.\ \ref{MR}) associated with each
  EoS. The quantity $\chi$ denotes the fraction of quark matter inside
  of the most massive neutron star for each EoS.}
\label{fig:EoS}
\end{figure}
The condition of global electric charge
neutrality is given by the equation
\begin{equation}
(1 - \chi) \sum_{i=B,l} q_i^H \, n_i^H + \chi \sum_{i=q,l}q_i^q \,
  n_i^q = 0 \, ,
\end{equation}
where $q_i$ is the electric charge of particle species $i$, expressed
in units of the electron charge.  Because of the global conservation
of electric charge and baryonic number, the pressure in the mixed
phase increases monotonically with increasing energy density, as shown in
Fig.\ \ref{fig:EoS}. In this work we have chosen global rather than local
electric charge neutrality, since the latter is not fully consistent
with the Einstein-Maxwell equations and the micro-physical condition
of chemical equilibrium and relativistic quantum statistics, as shown
in \cite{Rufini2011}.  In contrast to local electric charge
neutrality, the global neutrality condition puts a net positive
electric charge on hadronic matter, rendering it more isospin
symmetric, and a net negative electric charge on quark matter,
allowing neutron star matter to settle down in a lower energy state
that otherwise possible \cite{glendenning92:a,glendenning01:a}.

\section{Structure of neutron stars}
\label{sect:5}

\begin{figure}[!hbtp]
\includegraphics[width=1.0\linewidth,clip]{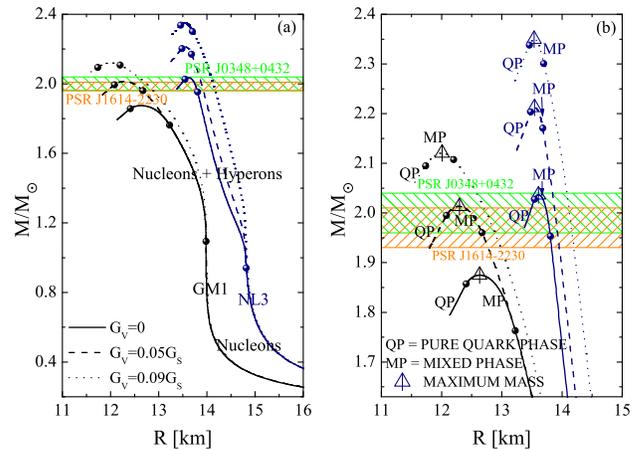}
\caption{ (Color online) (a) Mass-radius relationships of
    neutron stars made of quark-hybrid matter.  (b) Enlargement of the
    maximum mass region of (a). The labels GM1 and NL3 label the
    hadronic model for the EoS, $G_V$ indicates the strength of the
    vector coupling constant among quarks.}
\label{MR}
\end{figure}

We now turn to the discussion of the structure of neutron
stars, which are computed for the microscopic models of quark-hybrid
matter discussed in Sect.\ \ref{sect:2} and \ref{sect:3}.  We shall
explore three values for the vector coupling constant $G_V/G_S$, i.e.\
 0,  0.05, and 0.09. The mass-radius relationships of these
stars are shown in Fig.~\ref{MR}.
One sees that increasing values of $G_V$ lead to higher maximum masses
for both equations of state (GM1 and NL3) studied in this work. This
is expected, since the stiffness of the equation of state increases
with $G_V$. We also note that all neutron stars computed for NL3 have
larger radii than those obtained for the GM1 parametrization. This is
so because the NL3 equation of state is stiffer than the GM1 equation
of state, leading to quark deconfinement at densities that are lower
than for the GM1 parametrization, as can be seen in
Fig.\ (\ref{fig:EoS}).  This figure also shows that the neutron stars
close to the mass peaks possess extended mixed phase regions with
approximately 40\% quark matter for NL3 and 35\% quark matter for
GM1. In addition, we find that calculations carried out for
GM1 and a vanishingly small value of $G_V$ lead to neutron star masses
of around $1.8 \, M_\odot$, which is well below the masses observed
for neutron stars J1614-2230 ($1.97\pm 0.04\, \msun$)
\citep{Demorest2010} and J0348+0432 ($2.01 \pm 0.04 \,\msun$)
\citep{Antoniadis13}. Therefore, this combination of model
parameters for the equation of state can be ruled out. The
  calculations have been carried out using a combination of the
  Baym-Pethick Sutherland \citep{BPS} and Baym-Bethe-Pethick
  \citep{BBP} EoS at sub-saturation densities.

\section{Thermal Evolution}
\label{sect:6}

We now turn our attention to the thermal evolution of neutron stars
whose structure and interior composition are given by the microscopic
models described in the previous sections.

Here we briefly describe the thermal evolution equations that govern
the cooling of neutron stars. The thermal balance and transport
equations for a general relativistic, spherically symmetric object is
given by
\begin{align}
 \frac{\partial{(L e^{\nu})}}{\partial{r}}&=-\frac{4\pi
   r^2}{\sqrt{1-2m/r}} \left[\epsilon_\nu e^{\nu} + c_v
   \frac{\partial{(T e^{\nu /2})}}{\partial{t}} \right],
\label{eq:enbal}
\\ \frac{L e^{\nu}}{4 \pi r^2 \kappa}&=\sqrt{1-2m/r} \,
\frac{\partial{(T e^{\nu/2})}}{\partial{r}}\, ,
\label{eq:entransp}
\end{align}
where $r$, $m(r)$, $\rho(r)$, and $\nu(r)$ represent the radial
distance, mass, energy density, and gravitational potential,
respectively. Furthermore, the thermal variables are given by the
interior temperature $T(r,t)$, the luminosity $L(r,t)$, neutrino
emissivity $\epsilon_\nu(r, T )$, thermal conductivity $\kappa(r,T)$,
and the specific heat per unit volume $c_v(r,T)$.

The solution of Eqs.~(\ref{eq:enbal})--(\ref{eq:entransp}) is obtained
with the help of two boundary conditions, one at the core, where the
luminosity vanishes, $L(r=0)=0$, since the heat flux there is
zero. The second boundary condition has to do with the surface
luminosity, which is defined by the relationship between the mantle
temperature and the surface temperature
\citep{2004ApJS..155..623P,Yakovlev2004,Page2009}. Furthermore we consider all neutrino emission processes relevant to the thermal evolution of compact stars, including the Pair Breaking and Formation process (PBF) responsible for a splash of neutrinos on the onset of pair formation. We now describe
pairing effects and its corresponding effects on the thermal evolution
on the quark-hybrid stars discussed in this paper.

\subsection{Pairing Models}

In addition to the microscopic model described in the previous
sections, we now devote some time to the discussion of pairing of
nucleons, which will be used in our cooling simulations below. Pairing
among nucleons has received enormous interest recently due to the
unusual thermal data observed for the neutron star in Cassiopeia A
(Cas A) (see e.g.\ \citep{Page2011,Yakovlev2011,blaschke12:a} for a
recent study of the effects of pairing in the thermal evolution of
compact stars).

A full-blown microscopic description of pairing among neutrons and
protons in beta stable matter at high densities is a challenging task,
and so far there are still many uncertainties, particularly with
respect to proton-pairing at high-densities \cite{Chen1993}. In this
work we use a phenomenological description, as described in
\cite{2001PhR...354....1Y}. We assume that neutrons form singlet
$^1S_0$ pairs in the crust and triplet $^3P_2$ pairs in the core.

As for the protons, there is still much discussion as to how high densities
protons may form pairs in the cores of neutron stars.  As discussed in
\cite{Steiner2006}, the presence of the direct Urca process in the
core of neutron stars depends strongly on the symmetry energy and on
its possible dependence on a so-called ``quartic term" (a term that is
of fourth order in the deviation from symmetric matter). As pointed
out in \cite{Steiner2006}, it is very difficult to interpret neutron
star cooling without more information regarding the symmetry energy
and its ``quartic term" dependence at high densities. For that reason
we have chosen to study three different models for proton pairing,
\begin{figure}[!hbtp]
\includegraphics[width=1.0\linewidth,clip]{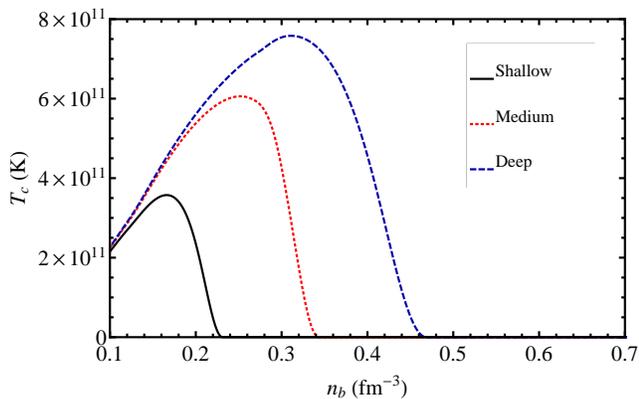}
\caption{(Color online) Critical temperature, $T_C$, for the
    onset of proton ($^1S_0$) singlet pairing in neutron star matter
    as a function of baryon number density, $n_b$.}\label{fig:3patt}
\end{figure}
which are referred to as shallow, medium, and deep. As the labeling
indicates, these models correspond to proton pairing that ends at low,
medium and high densities, respectively.  In Fig.~\ref{fig:3patt} we
show the critical temperature for the three models used for proton
  $^1S_0$ singlet pairing in the cores of the stars studied in this
  paper.

\subsection{Neutron Star Cooling Curves}

We now show the thermal evolution obtained by numerically integrating
the energy balance and transport equations
(\ref{eq:enbal})--(\ref{eq:entransp}).  We have chosen to perform
simulations on two different neutron stars; the first has a
gravitational mass of $1.4\, M_\odot$, and the second has a mass that
is closer to the maximum-mass value of each stellar sequence.  The
thermal evolution of neutron stars with masses between these two
limiting
\begin{table}
\begin{center}
  \caption{Properties of the neutron stars whose thermal
        evolutions are investigated for the nuclear parametrizations
        (GM1 and NL3) of this paper. $G_V$ is the vector coupling
        constant among quarks, $M$ and $R$ denote the neutron stars'
        gravitational masses and radii, respectively.}
  \label{tab:1}
\begin{tabular}{l c c c l}
\hline
\hline
    Parametrization &  $G_V/G_S$ & $M$ $[M_\odot]$ & $R$ [km] &
        $\rho_c$ [MeV/fm$^{3}$]  \\
\hline
GM1 & 0     & 1.4   & 13.84  &  375.27      \\
    &       & 1.87  & 12.63  &  961.10    \\
\hline
GM1 & 0.05  & 1.4   & 13.85  &  375.27          \\
    &       & 2.0   & 12.48  &  819.58     \\
\hline
GM1 & 0.09  & 1.4   & 13.91 &   355.08          \\
    &       & 2.0   & 12.74  &  443.28     \\
\hline
NL3 & 0     & 1.4   & 14.32  &  343.30          \\
    &       & 2.0   & 13.75  &  541.10      \\
\hline
NL3 & 0.05  & 1.4   & 14.47  &  333.76        \\
    &       & 2.2   & 13.62  &  675.07     \\
\hline
NL3 & 0.09  & 1.4   & 14.68  &  311.33            \\
    &       & 2.0   & 14.12  &  693.88     \\
\hline
\hline
\end{tabular}
\end{center}
\end{table}
cases will then lie within the bounds of these two cooling curves.
  In Table \ref{tab:1} we show the properties of the neutron stars
  whose thermal evolution is being studied.

\begin{figure}[!hbtp]
\includegraphics[width=1.0\linewidth,clip]{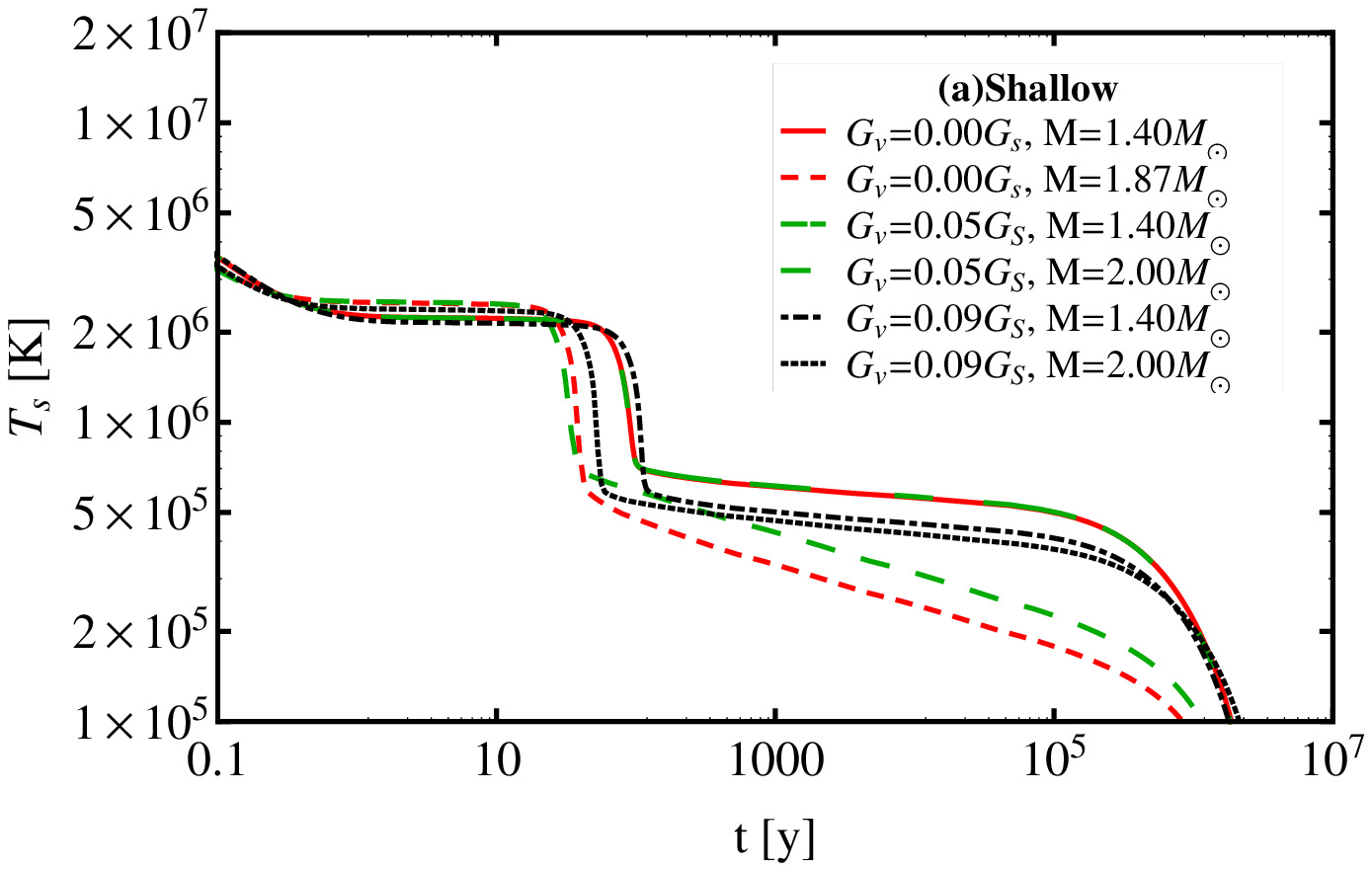}
\includegraphics[width=1.0\linewidth,clip]{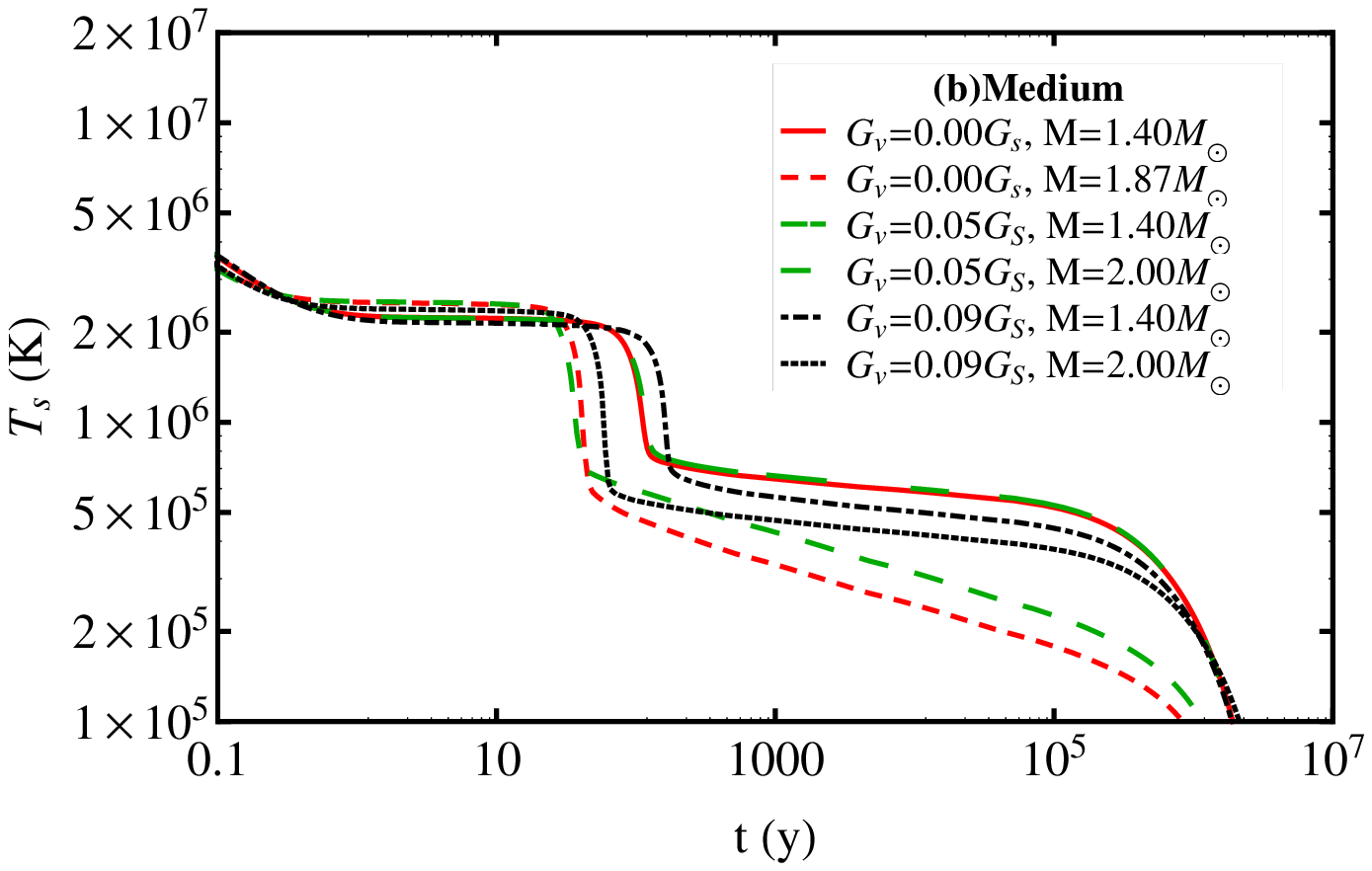}
\includegraphics[width=1.0\linewidth,clip]{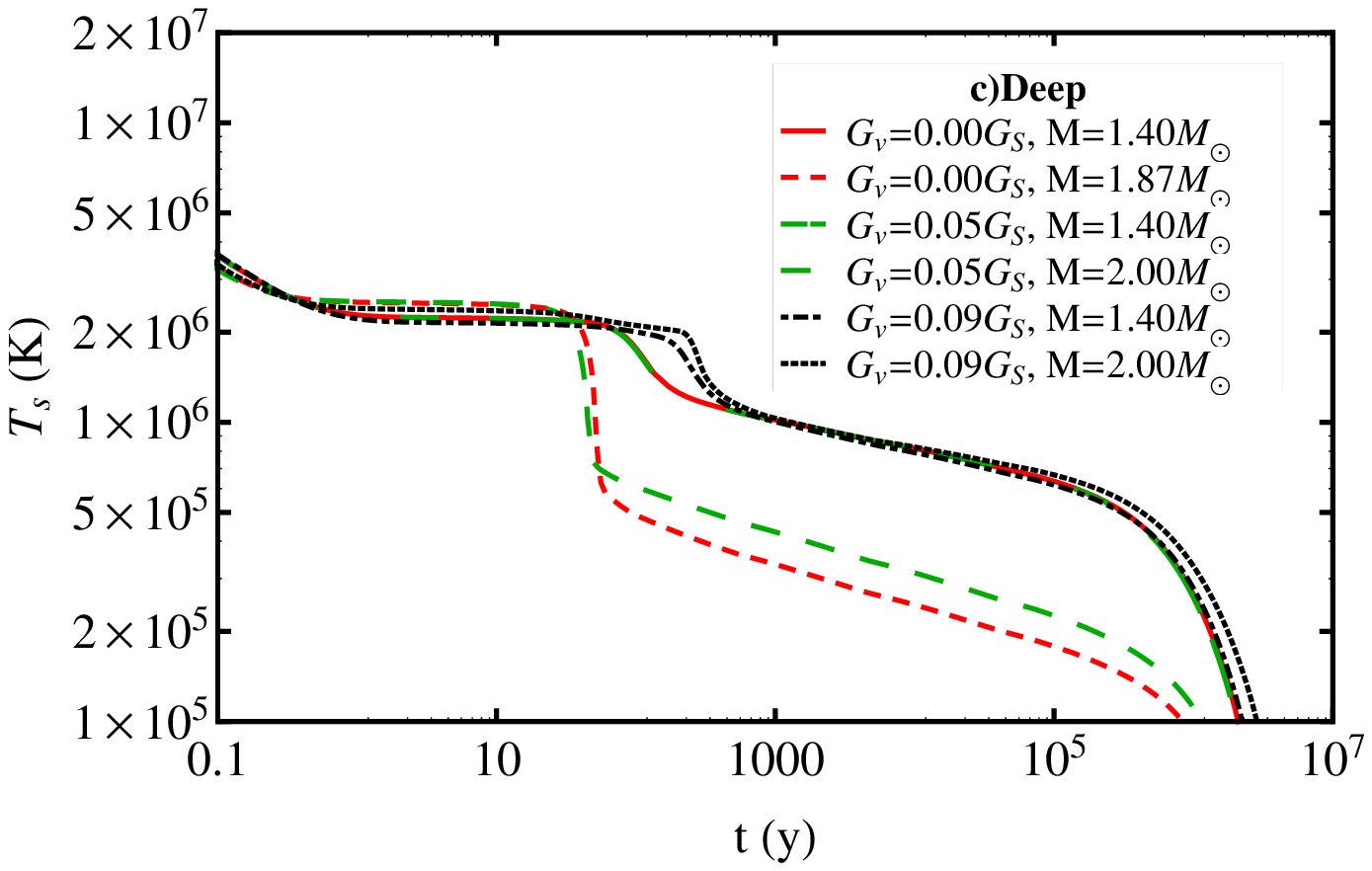}
\caption{(Color online) Thermal evolution of neutron stars for the
  different proton pairing scenarios (shallow, medium, and deep)
  considered in this paper. The calculations are carried out for the
  GM1 parametrization and vector coupling constants ranging from zero
  to $0.09\, G_S$.}\label{fig:res1}
\end{figure}

\begin{figure}
\includegraphics[width=1.0\linewidth,clip]{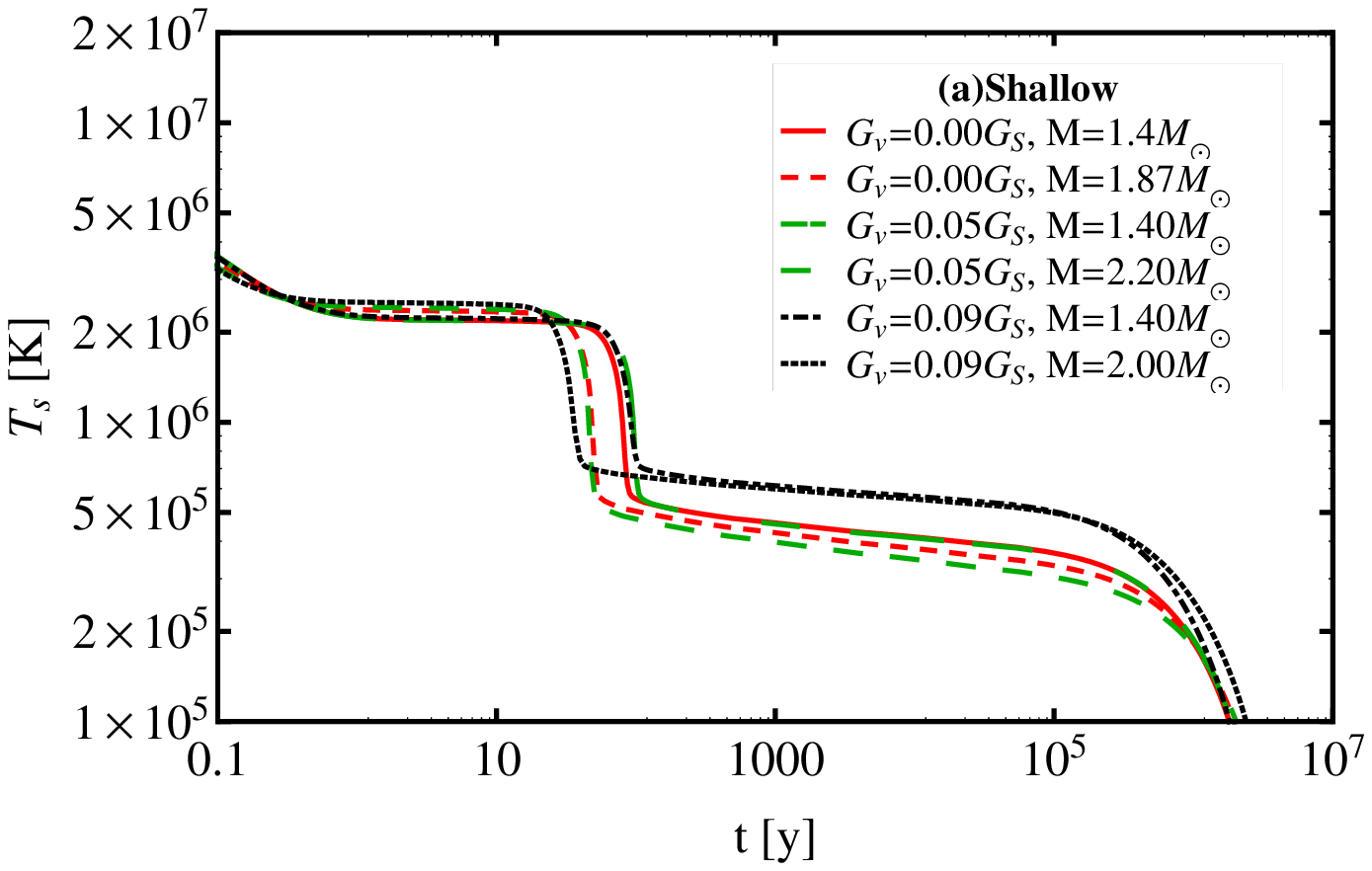}
\includegraphics[width=1.0\linewidth,clip]{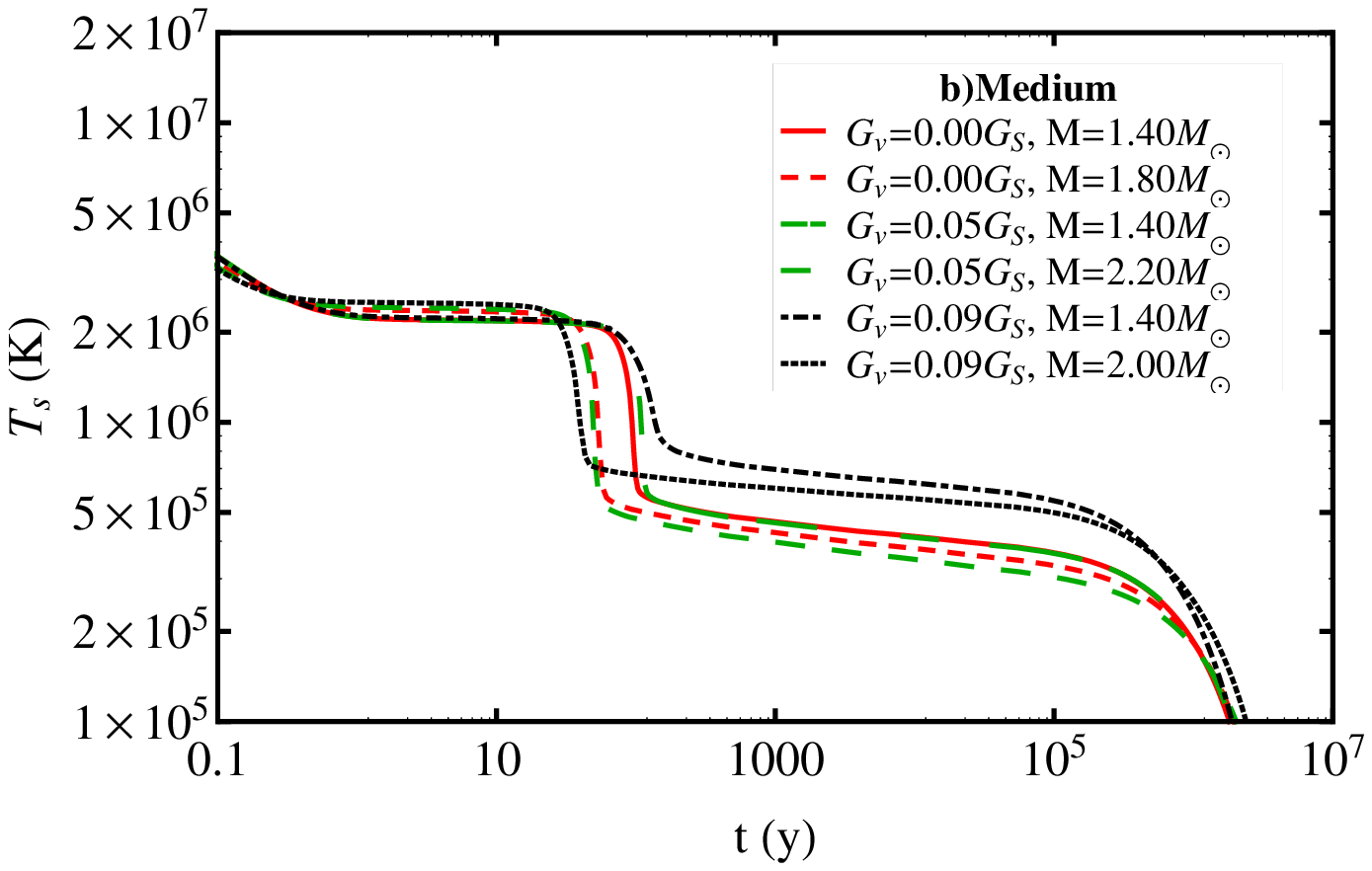}
\includegraphics[width=1.0\linewidth,clip]{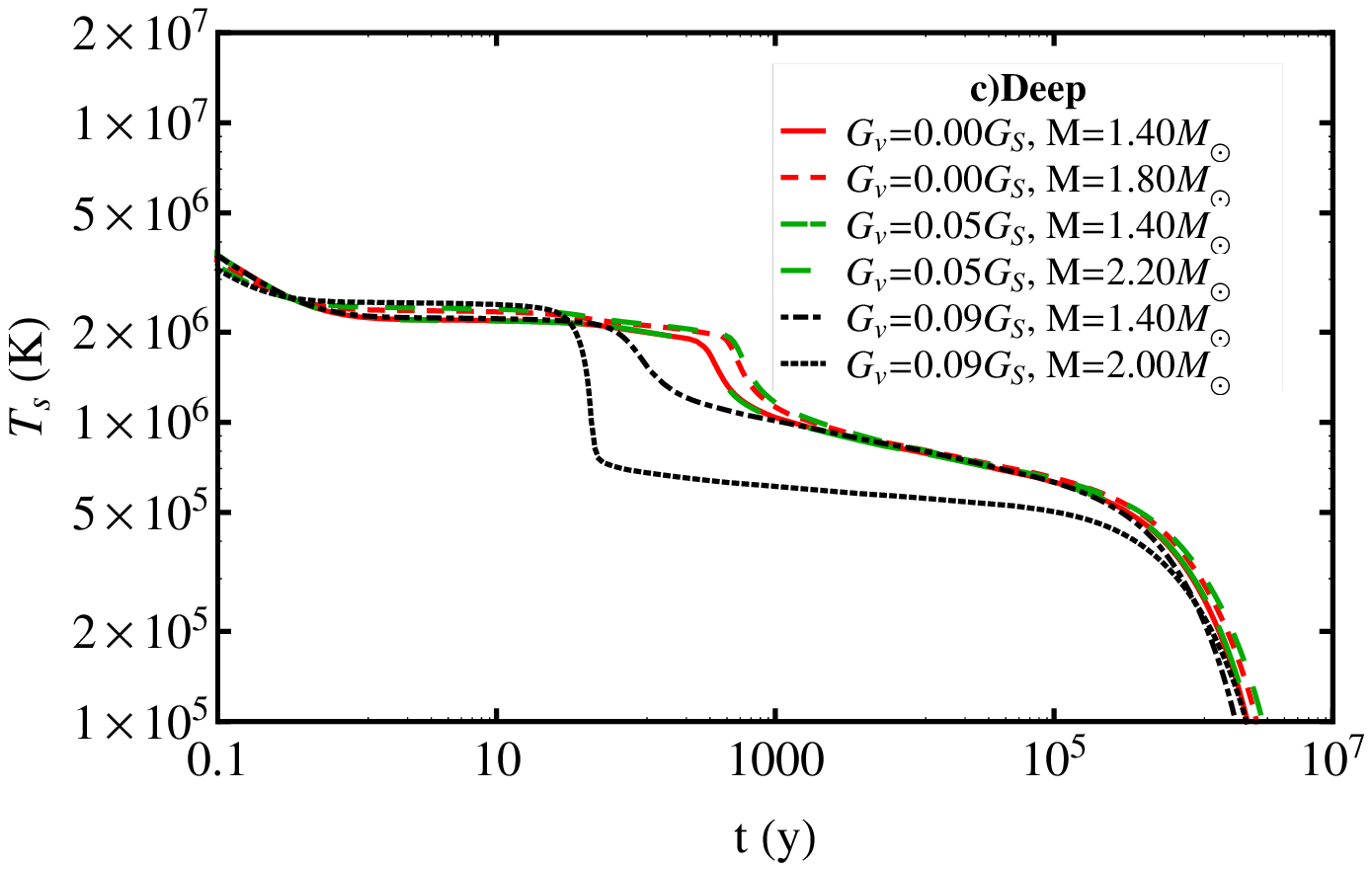}
\caption{(Color online) Same as Fig.\ \ref{fig:res1}, but for the NL3
  parametrization.}\label{fig:res2}
\end{figure}

We show in Fig.~\ref{fig:res1} the surface temperature $T_s$ as a
function of time $t$ (in years) for the GM1 parametrization. The
results for the NL3 parametrization are shown in
Fig.~\ref{fig:res2}. Figure \ref{fig:res1} shows that the shallow and
medium proton pairing cases obtained for the GM1 parametrization
exhibit very little difference. For both cases pairing is not strong
enough to completely suppress all fast neutrino processes so that
these stars exhibit fast cooling. Stars with a lower mass show
slightly slower cooling due to their smaller core densities (and thus
smaller proton fractions). The situation is different for the deep
pairing model where only for the lighter stars ($G_V = 0$ and $G_V =
0.05 \, G_S$) the neutrino process is completely suppressed.
Furthermore, for the deep pairing model, all stars obtained for $G_V =
0.09 \, G_S$ have their fast neutrino processes suppressed.

As for the NL3 parametrization, Fig.~\ref{fig:res2} shows that the
shallow and medium cases have similar behavior in that there is no
complete suppression of the fast neutrino processes. The deep case,
however, exhibits a behavior that is somewhat opposite to what we
found for the GM1 parametrization.  One sees that the fast neutrino
processes are totally suppressed in both light and heavy neutron stars
when $G_V = 0$ and $G_V = 0.05 \, G_S$.  In contrast to this, when
$G_V = 0.09 \, G_S$ we see that the fast neutrino processes are only
suppressed in lighter neutron stars.


\subsection{Comparison with Observed Data}

We complete our study of the thermal evolution by comparing our
results with the thermal behavior observed for compact stars,
and, in particular, that of the neutron stars in Cassiopeia A (Cas
A). This object is the youngest known neutron star from which the
thermal emission has been observed continuously for a decade.  Heinke
\& Ho found that the surface temperature of Cas A has dropped by $4\%$
between $2000$ and $2009$, from $2.12$ to $2.04\times10^6$~K
\cite{HeinkeHo2010}. The rapid cooling has been attributed to the
onset of neutron superfluidity in the stellar core. The observed data
for the neutron star in Cas A has been revisited recently by Ho {\it
  et al.} (2015) \cite{2015PhRvC..91a5806H}, where two new Chandra
ACIS-S graded observations are presented. We note, however, that the 
statistical significance of Cas A observed data has been called
into question, as discussed in Ref. \cite{Posselt:2013xva}.

\begin{figure}
\includegraphics[width=0.9\linewidth,clip]{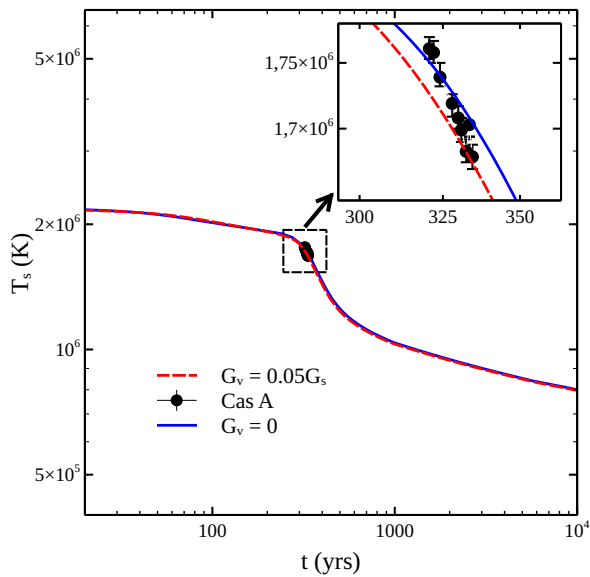}
\caption{(Color online) Cooling curves of a $M=1.4\, M_\odot$
      neutron star computed for NL3 and vector coupling constants
      $G_V=0$ and $G_V=0.05 \, G_S$. The inset shows the data observed
      for Cas A over a time period of one decade. (Data taken from
      \citep{2015PhRvC..91a5806H}.)} \label{fig:res3}
\end{figure}

For the models of this paper, only the $1.4\, M_\odot$ neutron star
computed for the NL3 parametrization ($G_V=0$ and $G_V=0.05 G_S$) and
with deep pairing for the protons agrees with the data observed for
Cas A. The reason being that this model strongly suppresses the fast
neutrino cooling processes, which is necessary to explain the thermal
behavior of Cas A as discussed in \cite{Page2011,Yakovlev2011}. We
show this result in Fig.~\ref{fig:res3}.
We note that agreement with the observed Cas A data is obtained
  for neutron stars with masses of $1.4\, M_\odot$, before the onset of quark
  matter in the stellar core. This result is in agreement with
  estimates of the mass of Cas A \cite{2015PhRvC..91a5806H}, which indicates
  that the mass of this object is probably too low to contain quark
  matter.

 We now confront the two NL3 models that are in agreement with Cas A
 (i.e., $G_V=0$ and $G_V=0.05 G_S$ with deep proton pairing) with
 observed thermal data on compact stars. We use two sets of observed
 data (see \cite{Page2009} and references therein), one for age
 estimates based on the stars' spin-down properties, and the other
 based on the so called kinematic age. We note that the kinematic ages
 constitue more realistic age estimates as they are associated with
 kinematic properties of supernovae belived to be the progenitors of
 the neutron stars in question. For the few cases where both kinematic
 and spin-down ages have been estimated, large discrepancies have been
 found. This indicates that the spin-down age needs to be considered
 very carefuly, perhaps serving only as an upper limit on the true
 age of a given neutron star.

\begin{figure}
\includegraphics[width=0.9\linewidth,clip]{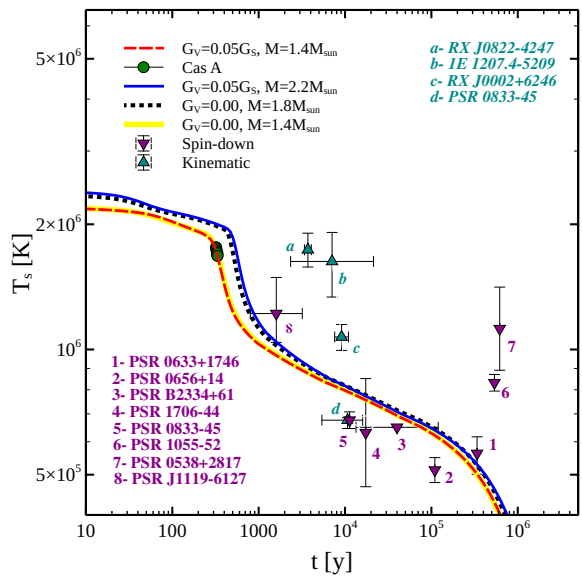}
\caption{(Color online) Theoretical cooling curves of neutron
    stars, computed for NL3 and vector coupling constants $G_V=0$ and
    $G_V=0.05 \, G_S$, compared with observed data.  Pink (green) diamonds denote
    spin-down (kinematic) age estimates.} \label{fig:cooling_obs}
\end{figure}

In Fig.~\ref{fig:cooling_obs}, we compare the cooling tracks of neutron
stars computed for the NL3 parametrization with the observed data.
We note that our model agrees fairly well with some observed data, but
fails to reproduce the data of several other neutron stars. This is
not an inherent feature of our model but, rather, appears symptomatic for
most thermal models that agree with the Cas A data (see for instance \cite{Yakovlev2011}). It may indicate that these objects are subjected
to a heating mechanism which keeps them warm during their evolution.

\section{Summary and conclusions}
\label{sect:7}

In this work, we have used an extension of the non-local 3-flavor
Nambu-Jona Lasinio model to study quark deconfinement in the cores of
neutron stars. Confined hadronic matter is described by non-linear
relativistic nuclear field theory, adopting two popular hadronic
parametrizations labeled GM1 and NL3.  The phase transition from
confined hadronic matter to deconfined quark matter is modeled via the
Gibbs condition, imposing global electric charge neutrality on the
particle composition of neutron star matter. Repulsive forces among
the quarks are described in terms of a vector coupling constant,
$G_V$, whose value ranges from zero (no repulsion) to $0.9\, G_S$,
where $G_S$ denotes the scalar strong coupling constant of the theory.

Each one of our models for the EoS of (quark-hybrid) neutron star
matter accommodates high-mass neutron stars with masses up to 2.4
solar masses as long as the value of $G_V$ is sufficiently large.  All
high-mass stars contain extended quark-hybrid matter cores in their
centers, but a pure quark matter is never reached for any of our model
parametrizations. The maximum neutron star masses drop if the strength
of the vector repulsion among quarks is reduced, falling below the
$2\, M_\odot$ limit set by pulsars $J1614-2230$ ($1.97 \pm 0.04 \,
M_{\odot}$)~\cite{Demorest2010} and $J0348+0432$ ($2.01 \pm 0.04\,
M_{\odot}$)~\cite{Antoniadis13} for some parameter combinations.
Examples of this are neutron stars computed for the GM1
parametrization with $G_V=0$, which yields a maximum-mass neutron star
of $\sim 1.8 \, M_\odot$, in conflict with the recent mass
determinations mentioned just above.

The quark-hybrid stars of our study possess relatively high proton
fractions in their cores so that fast neutrino processes, most notably
the direct Urca process, is active, leading to very rapid stellar
cooling. An agreement with the thermal evolution data of the neutron
star in Cas A can be obtained, however, if one assumes that strong
proton-pairing is occurring in the core of this neutron star, which is
known to strongly suppress fast cooling. Under this condition, a
$1.4\, M_\odot$ neutron star computed for the NL3 model and values of
the vector coupling constants between $G_V=0$ and $G_V=0.05\, G_S$
lead to excellent agreement with the observed data.  It is important
to note that the proton pairing model has not been fine tuned to the
Cas A data. Further studies where we also take into account pairing
among the quarks will be presented in a future work. We have
  also compared the thermal behavior predicted for Cas A with that of
  other compact stars. In contrast to Cas A, however, these
  observations only temperature "snapshots" exist for the latter, with
  age estimates based on either their kinematic or spin-down
  properties. The results show that the NJL models that are in
  agreement with Cas A (i.e., $G_V=0$ and $G_V=0.05 G_S$ with deep
  proton pairing) lead to good agreement with the observed data of
  several other compact objets, while failing to reproduce seveal
  other data.  This, however, is not an inherent feature of our model but,
  rather, appears symptomatic for most thermal models that agree with
  the Cas A data \cite{Yakovlev2011}. 

\section*{Acknowledgments}

S.M.C.\ acknowledge the support by the International Cooperation
Program CAPES-ICRANet financed by CAPES -- Brazilian Federal Agency
for Support and Evaluation of Graduate Education within the Ministry
of Education of Brazil. R. N.\ acknowledges financial support by CAPES
and CNPq. M.O.\ and G.C.\ acknowledge financial support by CONICET,
Argentina.  F.W.\ is supported by the National Science Foundation
(USA) under Grant PHY-1411708.

\end{document}